\documentclass[12pt]{article}
\usepackage[left=2.54cm,top=2.54cm,right=2.54cm,bottom=2.54cm]{geometry}
\usepackage{amsmath, amssymb}
\usepackage[latin1]{inputenc}
\usepackage{mathrsfs}
\usepackage{cite}
\usepackage{dsfont}

\begin{document}
\date{}
\title{Exact Solutions of the 2D Dunkl--Klein--Gordon Equation: The Coulomb Potential and the Klein--Gordon Oscillator}

\author{R. D. Mota$^{a}$, D. Ojeda-Guill\'en$^{b}$\footnote{{\it E-mail address:} dojedag@ipn.mx}, \\ M. Salazar-Ram\'{\i}rez$^{b}$, and
 V. D. Granados$^{c}$} \maketitle

\begin{minipage}{0.9\textwidth}

\small $^{a}$ Escuela Superior de Ingenier{\'i}a Mec\'anica y El\'ectrica, Unidad Culhuac\'an,
Instituto Polit\'ecnico Nacional, Av. Santa Ana No. 1000, Col. San
Francisco Culhuac\'an, Del. Coyoac\'an, C.P. 04430, Ciudad de M\'exico, Mexico.\\

\small $^{b}$ Escuela Superior de C\'omputo, Instituto Polit\'ecnico Nacional,
Av. Juan de Dios B\'atiz esq. Av. Miguel Oth\'on de Mendiz\'abal, Col. Lindavista,
Del. Gustavo A. Madero, C.P. 07738, Ciudad de M\'exico, Mexico.\\

\small $^{c}$ Escuela Superior de F{\'i}sica y Matem\'aticas,
Instituto Polit\'ecnico Nacional, Ed. 9, Unidad Profesional Adolfo L\'opez Mateos, Del. Gustavo A. Madero, C.P. 07738, Ciudad de M\'exico, Mexico.\\

\end{minipage}

\begin{abstract}
We introduce the Dunkl--Klein--Gordon (DKG) equation in 2D by changing the standard partial derivatives by the Dunkl derivatives in the standard Klein--Gordon (KG) equation. We show that the generalization with Dunkl derivative of the $z$-component of the angular momentum is what allows the separation of variables of the DKG equation. Then, we compute the energy spectrum and eigenfunctions of the DKG equations for the 2D Coulomb potential and the Klein--Gordon oscillator analytically and from an ${\rm su}(1,1)$ algebraic point of view. Finally, we show that if the parameters of the Dunkl derivative vanish, the obtained results suitably reduce to those reported in the literature for these 2D problems.
\end{abstract}


\section{Introduction}

The reflection operators were introduced by Wigner \cite{wigner} to generalize the boson quantization rules and were applied by Yang to the deformed one-dimensional harmonic oscillators \cite{yang}. The reflection operators are also closely related to the solutions of the Calogero's quantum mechanical models \cite{br,ply} and the integrable models of quantum mechanics, as the Calogero--Sutherland--Moser models  \cite{hikami,kakei,lapon}.

The Dunkl operators are combinations of differential and difference operators and are associated to a finite reflection group, also called finite Coxeter group. Dunkl reintroduced these operators to study polynomials with discrete symmetry groups in several variables  \cite{dunkl1}. The Dunkl operators have been applied to the study of Laplace operators in $\bf  R^n$. It has been shown that the Dunkl--Laplace operator can be written as the classical Laplacian plus other terms that depend on the reflection operator, such that the resulting operator is no longer invariant under the whole orthogonal group $\mathcal{O}(m)$ but only under a finite reflection group $\mathcal{G}$, which is a finite subgroup of $\mathcal{O}(m)$ \cite{dunkl1,dunkl2,Bie}.

The Dunkl derivatives, closely related to the so-called Bannai--Ito and Dunkl--Schwinger algebras, have been applied to solve the Schr\"odinger equation for the Coulomb problem and the harmonic oscillator in two and three dimensions. It has been shown that the Jacobi, Legendre, Hermite and -1 polynomials are involved in solving these problems \cite{GEN1,GEN2,GEN3,GEN4}. Also, in Refs. \cite{nos1,nos2} we used the Lie algebra ${\rm su}(1,1)$ and its irreducible representations to solved the two-dimensional Dunkl-oscillator and the Dunkl--Coulomb problems. Recently, the Schr\"odinger equation for the Dunkl--Coulomb problem in 3D has been solved and its superintegrability and dynamical symmetry have been studied \cite{sami1,sami2}. In the relativistic regime, in Ref. \cite{nos3} we studied the problem of the Dirac--Dunkl oscillator in two dimensions.

In this paper we study the Dunkl--Klein--Gordon DKG equation for the Coulomb potential and the Klein--Gordon oscillator in two dimensions. What we call the DKG equation is the standard Klein--Gordon equation, in which the partial derivatives are changed for the Dunkl derivatives. The main purpose of the present paper is to show that DKG equation for the 2D Coulomb potential and the Klein--Gordon oscillator is exactly solvable and obtain their respective energy spectrum and eigenfunctions.
For each of these problems, we find the energy spectrum and eigenfunctions by introducing suitable sets of operators that span the algebra ${\rm su}(1,1)$ and using the theory of unitary representations. Also, we solve analytically the DKG equation for both of these problems and the energy spectrum and the eigenfunctions are found again.

This work is organized as follows. In section $2$, we obtain the DKG equation in 2D for the Coulomb potential. We show that the generalization with the Dunkl derivative of the $z$-component of the angular momentum is the one that allows the separation of variables for the DKG equation. By introducing a set of operators that span the Lie algebra ${\rm su}(1,1)$ and the tilting transformation, we obtain the energy spectrum and eigenfunctions from an algebraic point of view. Also, we find the energy spectrum and eigenfunctions of the DKG equation for the Coulomb potential analytically. In section $3$, we study the DKG equation in 2D for the Klein--Gordon oscillator. We follow the same procedure that we used for the Coulomb potential. Finally, in section $4$, we give some concluding remarks. It is argued that our results when the parameters of the Dunkl derivative vanish are in complete agreement with those reported in the literature for these 2D problems.

\section{The DKG equation for the 2D Coulomb potential}

The standard Klein--Gordon equation is given by
\begin{equation}
(\partial^\mu \partial_\mu +m^2c^4)\Psi_{C}=0.
\end{equation}
In two dimensions and for stationary states this equation takes the form
\begin{equation}
\left(E-V_c(\rho)\right)^2\Psi_{C}=\left(c^2\left({\bf P}+\frac{e}{c}{\bf A}\right)^2+(m+V_s(\rho))^2c^4\right)\Psi_{C},
\end{equation}
where we have assumed that a charged particle is subject to move under the influence of an electromagnetic field with vector potential $\bf A$, a vector potential $V_c(\rho)$, and a scalar potential $V_s(\rho)$. Thus, in polar coordinates $\rho=\sqrt{x^2+y^2}$, $\tan{\phi}=\frac{y}{x}$ for the problem with the Coulomb potential we have $V_c(\rho)=-\frac{Ze^2}{\rho}$, $A_1=A_2=0$ and $V_s=0$.

In Cartesian coordinates, if we change in the standard Klein--Gordon equation the partial derivatives $\frac{\partial}{\partial x}$ and $\frac{\partial}{\partial y}$ by the Dunkl derivatives
\begin{equation}
D_1\equiv\frac{\partial}{\partial x}+\frac{\mu_1}{x}(1-R_1), \quad\quad D_2\equiv\frac{\partial}{\partial y}+\frac{\mu_2}{y}(1-R_2),
\end{equation}
we obtain the DKG equation. In this definition, the constants satisfy $\mu_1>0$ and $\mu_2>0$ \cite{GEN4}, and $R_1$ and $R_2$ are the reflection operators with respect to the $x$- and $y$-coordinates, it is to say, $R_1f(x,y)=f(-x,y)$ and $R_2f(x,y)=f(x,-y)$. Therefore, ${\bf P}^2=-\hbar^2 \nabla^2$ changes to
${\bf P}^2=-\hbar^2\left(D_1^2+D_2^2\right)\equiv-\hbar^2 \nabla^2_D$, where  $ \nabla^2_D$ is known as the Dunkl Laplacian. Hence, the DKG equation to be solved in this section is
\begin{equation}
H_C\Psi_{C}\equiv\left(-\hbar^2c^2(D_1^2+D_2^2)+m^2c^4-\left(E-V_c\right)^2\right)\Psi_{C}=0,\hspace{2ex}\label{DKGC}
\end{equation}
where we have defined the pseudo-Hamiltonian $H_C$.

By using the action of the reflection operator $R_i$ on a function of two variables $f(x,y)$ it is obvious that
\begin{equation}
R_1D_1=-D_1R_1,\hspace{3ex} R_1^2=1, \hspace{3ex}\frac{\partial}{\partial x}R_{1}=-R_1\frac{\partial}{\partial x}, \hspace{3ex}R_1x=-xR_1.\label{pro1}
\end{equation}
Operators associated with $y$-direction have similar properties. Also, we show that the following equalities hold
\begin{equation}
R_1R_2=R_2R_1,\hspace{2ex}[D_1,D_2]=0,\hspace{2ex}[x_i,D_j]=\delta_{ij}+2\mu_{\delta{ij}}R_{\delta{ij}}\hspace{1ex}\hbox{(no sum over $i$ and $j$)}.\label{pro2}
\end{equation}

At this stage it is convenient to introduce the $z$-component of the Dunkl angular momentum $L_z=-i\hbar(xD_2-yD_1)$. Thus, we show the following results
\begin{eqnarray}
&&\left[xD_2,D_1^2+D_2^2\right]=2D_2D_1,\label{res1}\\
&&\left[yD_1,D_1^2+D_2^2\right]=2D_1D_2,\label{res2}\\
&&\frac{\mu_i}{x_i}\left(1-R_i\right)V(\rho)=V(\rho)\frac{\mu_i}{x_i}\left(1-R_i\right), \hspace{2ex}i=1,2,\label{res3}\\
&&\left(x\frac{\partial}{\partial y}-y\frac{\partial}{\partial x}\right)V(\rho)=\frac{\partial }{\partial \phi}V(\rho)=V(\rho)\frac{\partial }{\partial \phi}.\label{res4}
\end{eqnarray}
From these results it follows immediately that
\begin{equation}
[L_z, H_C]=0.
\end{equation}
This means that the operator $L_z$ is a constant of motion of the pseudo-Hamiltonian $H_C$. As it is shown below,
this fact allows us to solve the DKG equation (\ref{DKGC}) by using separation of variables on the DKG wave function.

Explicitly, the Dunkl Laplacian in Cartesian coordinates takes the form
\begin{eqnarray}
&&\nabla_D^2= D_1^2+D_2^2\\
&&\hspace{4ex}=\frac{\partial^2}{\partial x^2}+\frac{\partial^2}{\partial y^2}+2\frac{\mu_1}{x}\frac{\partial}{\partial x}+2\frac{\mu_2}{y}\frac{\partial}{\partial y}-\frac{\mu_1}{x^2}(1-R_1)-\frac{\mu_2}{y^2}(1-R_2),
\end{eqnarray}
or in  polar coordinates it is written as
\begin{equation}
\nabla_D^2= \frac{\partial^2}{\partial \rho^2}+\frac{1+2\mu_1+2\mu_2}{\rho}\frac{\partial}{\partial \rho}-\frac{2}{\rho^2}B_\phi, \label{laplapol}
\end{equation}
where $B_\phi$ is given by
\begin{equation}
B_\phi\equiv-\frac{1}{2}\frac{\partial^2}{\partial \phi^2}+\left(\mu_x\tan{\phi}-\mu_y\cot{\phi}\right)\frac{\partial}{\partial \phi}
+\frac{\mu_1 (1-R_1)}{2\cos^2{\phi}}+\frac{\mu_2 (1-R_2)}{2\sin^2{\phi}}.
\end{equation}
This operator is directly related to Dunkl orbital angular momentum $L_z$. In fact, it can be shown that
\begin{equation}
\frac{L_z^2}{\hbar^2}=2B_\phi+2\mu_1\mu_2(1-R_1R_2).
\end{equation}
Considering that $ [R_1,H_C]=[R_2,H_C]=0$, the eigenvalues and eigenfunctions of the angular part operator $B_\phi$ have been constructed in Refs. \cite{GEN1,GEN4}. If $\Phi(\phi)$ are the eigenfunctions and $\frac{s^2}{2}$ their corresponding eigenvalues, we have
\begin{equation}
B_\phi\Phi(\phi)=\frac{s^2}{2}\Phi(\phi)\label{esf}.
\end{equation}
Explicitly, the eigenfunctions $\Phi(\phi)$ are labeled in terms of the quantum numbers ($e_1,e_2$), which correspond to the eigenvalues ($1-2e_1,1-2e_2$) of the reflection operators ($R_1,R_2$), and are written in terms of the Jacobi polynomials $P_\ell^{(\alpha,\beta)} (x)$ as
\begin{equation}
\Phi_\ell^{(e_1,e_2)}(\phi)=\eta_\ell^{(e_1,e_2)}\cos^{e_1}\phi\sin^{e_2}\phi \hspace{0.2cm} P_{\ell-e_1/2-e_2/2}^{\mu_1-1/2+e_1,\mu_2-1/2+e_2}(-\cos{2\phi}).\label{angular}
\end{equation}
Here, the possible values that $\ell$ can take are restricted according to
\begin{equation*}
  (e_1,e_2)\in\left\lbrace
  \begin{array}{l}
      \left\{(0,0),(1,1)\right\},\hspace{1.0cm}\text{$\ell$ is a non-negative integer },\\
      \left\{(1,0),(0,1)\right\},\hspace{1.0cm}\text{$\ell$ is a positive half-integer }.\\
  \end{array}
  \right.
\end{equation*}
The factors $\eta_\ell^{(e_1,e_2)}$ are the normalization constants, and are given by
\begin{align}
\eta_\ell^{(e_1,e_2)}=&\sqrt{\left(\frac{2\ell+\mu_1+\mu_2}{2}\right)\left(\ell-\frac{e_1+e_2}{2}\right)}\nonumber\times\\
&\sqrt{\frac{\Gamma\left(\ell+\mu_1+\mu_2+\frac{e_1+e_2}{2}\right)}{\Gamma\left(\ell+\mu_1+\frac{1+e_1-e_2}{2}\right)\Gamma\left(\ell+\mu_2+\frac{1+e_2-e_1}{2}\right)}}.
\end{align}
Also, from these results the eigenvalues of equation (\ref{esf}) take the form
\begin{equation}\label{mcuad}
s^2=4\ell(\ell+\mu_1+\mu_2).
\end{equation}
From the orthogonality relation of the Jacobi polynomials, it can be deduced that the angular wavefunctions $\Phi_\ell^{(e_1,e_2)}(\phi)$ satisfy \cite{GEN1,GEN4}
\begin{equation}
\int_0^{2\pi}\Phi_\ell^{(e_1,e_2)}(\phi)\Phi_{\ell'}^{(e'_1,e'_2)}(\phi)|\cos{\phi}|^{2\mu_1}|\sin{\phi}|^{2\mu_2}d\phi=\delta_{\ell,\ell'}\delta_{e_1,e'_1}\delta_{e_2,e'_2}.
\end{equation}
In the rest of this section we concentrate on studying the radial part of the DKG equation for the Coulomb potential.

\subsection{Algebraic approach for the DKG--Coulomb problem}

Substituting the Laplacian operator in polar coordinates of Eqn. (\ref{laplapol}) into the expression (\ref{DKGC}), we obtain that the DKG equation takes the form
\begin{equation}
\left(E+\frac{Ze^2}{\rho}\right)^2\Psi_{C}=\left(-\hbar^2c^2\left(\frac{\partial^2}{\partial \rho^2}+\frac{1+2\mu_1+2\mu_2}{\rho}\frac{\partial}{\partial \rho}-\frac{2}{\rho^2}B_\phi\right)+m^2c^4\right)\Psi_{C}. \label{DKG}
\end{equation}
By setting the DKG wave function as $\Psi_{C}=R(\rho)\Phi(\phi)$, using the result of Eqn. (\ref{esf}), and the definitions
\begin{equation}
\lambda=\frac{2E\gamma}{\hbar c},\hspace{8ex}\gamma=\frac{Ze^2}{\hbar c}, \hspace{8ex}\alpha=\frac{m^2c^4-E^2}{\hbar^2c^2}\label{def},
\end{equation}
we can transform the Eqn. (\ref{DKG}) to the following radial equation
\begin{equation}
\left(\frac{d^2}{d \rho^2}+\frac{1+2\mu_1+2\mu_2}{\rho}\frac{d}{d \rho}-\frac{s^2-\gamma^2}{\rho^2}+\frac{\lambda}{\rho}-\alpha\right)R_{n\ell}(\rho)=0.\label{rad}
\end{equation}
Since we are interested in the bound states, from now on we suppose that $\alpha>0$.
In what follows we find the energy spectrum of the DKG equation by using the ${\rm su}(1,1)$ approach and the tilting transformation \cite{ADAMS}.

The Lie algebra ${\rm su}(1,1)$ is spanned by the generators $K_\pm=K_1\pm i K_2$, and $K_{0}$ , which satisfy the commutation relations \cite{vourdas}
\begin{eqnarray}
[K_{0},K_{\pm}]=\pm K_{\pm},\quad\quad [K_{-},K_{+}]=2K_{0}.\label{com}
\end{eqnarray}
The action of these operators on the Sturmian basis $\{|k,n\rangle, n=0,1,2,...\}$ is
\begin{eqnarray}
&&K_{+}|k,n\rangle=\sqrt{(n+1)(2k+n)}|k,n+1\rangle,\label{k+n}\\
&&K_{-}|k,n\rangle=\sqrt{n(2k+n-1)}|k,n-1\rangle,\label{k-n}\\
&&K_{0}|k,n\rangle=(k+n)|k,n\rangle,\label{k0n}
\end{eqnarray}
where $|k,0\rangle$ is the lowest normalized state. Equations (\ref{k+n})--(\ref{cas}) define the unitary irreducible representations
of the Lie algebra ${\rm su}(1,1)$. Thus, the number $k$ completely determines a representation of the algebra ${\rm su}(1,1)$. In the present work,
we restrict to the discrete series only, for which $k>0$. The Casimir operator for any irreducible representation of this algebra satisfies
\begin{equation}
K^{2}=K_0^2-K_1^2-K_2^2=-K_{+}K_{-}+K_{0}(K_{0}-1)=k(k-1).\label{cas}
\end{equation}

To find the energy spectrum for the DKG equation for the Coulomb potential, we introduce the set of operators
\begin{align}\label{A1}
A_0&=\frac{1}{2}\left(-\rho\frac{d^2}{d\rho^2}-2\left(1/2+\mu_1+\mu_2\right)\frac{d}{d\rho}+\frac{s^2-\gamma^2}{\rho}+\rho\right),\\\label{A2}
A_1&=\frac{1}{2}\left(-\rho\frac{d^2}{d\rho^2}-2\left(1/2+\mu_1+\mu_2\right)\frac{d}{d\rho}+\frac{s^2-\gamma^2}{\rho}-\rho
\right),\\\label{A3}
A_2&=-i\rho\left(\frac{d}{d\rho}+\frac{1}{\rho}\left(1/2+\mu_1+\mu_2\right)\right),
\end{align}
which span the Lie algebra ${\rm su}(1,1)$. These operators generalize those introduced by Barut
to study general central potentials \cite{AOB2,KTH}. A direct calculation shows that the Casimir operator for this algebra is
\begin{equation}
C^2=A_0^2-A_1^2-A_2^2=s^2-\gamma^2-\frac{1}{4}+(\mu_1+\mu_2)^2.
\end{equation}
According to Eqns. (\ref{mcuad}) and (\ref{cas}), the eigenvalues of $C^2$ must satisfy
\begin{equation}\label{opcas1}
4\ell(\ell+\mu_1+\mu_2)-\gamma^2-\frac{1}{4}+(\mu_1+\mu_2)^2=k(k-1).
\end{equation}
From this equation, we obtain that the group number $k$ (Bargmann index) for our problem is
\begin{equation}
k=\frac{1}{2} + \sqrt{(2\ell+\mu_1+\mu_2 )^2-\gamma^2},\label{k}
\end{equation}
where we have restricted to  $k>0$, since we are considering  the discrete series only.

The radial DKG equation (\ref{rad}) written in terms of the ${\rm su}(1,1)$ generators (\ref{A1}), (\ref{A2}) and (\ref{A3}) result to be
\begin{equation}
\left[-\left(A_0+A_1\right)+\lambda-\left(A_0-A_1\right)\alpha\right]R_{n\ell}(\rho)\equiv HR_{n\ell}(\rho)=0.\label{radial}
\end{equation}
Now, we introduce the similarity transformation
\begin{align}
&\widetilde{R}_{n\ell}(\rho)=e^{-i\theta A_2}R_{n\ell}(\rho), \label{eigen}\\
&\widetilde{H}=e^{-i\theta A_2}He^{i\theta A_2}.
\end{align}
The action of the scaling or tilting transformation on the operators $A_0$ and $A_1$ can be computed from the Baker--Campbell--Hausdorff formula. Thus,
\begin{align}
e^{-i\theta A_2}A_0e^{+i\theta A_2}&=A_0\cosh(\theta)+A_1\sinh(\theta),\\
e^{-i\theta A_2}A_1e^{+i\theta A_2}&=A_0\sinh(\theta)+A_1\cosh(\theta).
\end{align}
From these results it follows that
\begin{equation}
e^{-i\theta A_2}(A_0\pm A_1)e^{i\theta A_2}=e^{\pm\theta}(A_0\pm A_1).
\end{equation}
Thus, equation (\ref{radial}) can be written as
\begin{equation}\label{delt1}
\widetilde{H}\widetilde{R}(\rho)=\left[A_0 \left(-e^{\theta}-\alpha e^{-\theta}\right)
+ A_1 \left(-e^{\theta}+\alpha e^{-\theta}\right)+\lambda \right]\widetilde{R}_{n\ell}(\rho)=0.
\end{equation}
If we choose the scaling parameter as $\theta=\ln(\alpha)^{1/2}$, the coefficient of $A_1$ vanishes and therefore
\begin{equation}
\widetilde{H}\widetilde{R}(\rho)=\left[A_0 \left(-\alpha^{\frac{1}{2}}-\alpha^{\frac{1}{2}}\right)+\lambda\right]\widetilde{R}(\rho)=0,
\end{equation}
which implies
\begin{equation}
\left(A_0-\frac{\lambda}{2\sqrt{\alpha}}\right)\widetilde{R}(\rho)=0.
\end{equation}
Thus, from the action of the operator $A_0$ on the ${\rm su}(1,1)$ states (equation (\ref{k0n})) we obtain the relation
\begin{equation}
\frac{\lambda}{2\sqrt{\alpha}}=k+n.
\end{equation}
The energy spectrum of the DKG equation for the Coulomb potential can be obtained from this result by using the definitions of equation (\ref{def}), to obtain
\begin{equation}
E=\pm mc^2\left(1+\frac{\hbar c \gamma^2}{(n+k)^2} \right)^{-\frac{1}{2}}.
\end{equation}
Using the explicit form of the Bargmann index for our problem of equation (\ref{k}), we obtain that the energy spectrum of the DKG equation for the Coulomb potential is
\begin{equation}
E=\pm mc^2\left(1+\frac{\gamma^2}{(n+\sqrt{(2\ell+\mu_1+\mu_2 )^2-\gamma^2}+\frac{1}{2})^2} \right)^{-\frac{1}{2}}.\label{lieesp}
\end{equation}
It is well known that to solve analytically the Schr\"odinger and Klein--Gordon  radial equations for the Coulomb potential, it is necessary to introduce a new radial variable, which depends on the energy \cite{KTH}. However, our operators given in equations (\ref{A1})-(\ref{A3}) are free of a rescaling parameter, and therefore the group functions must also be rescaling-free. These group functions (necessary to apply the theory of  unitary representations) are known as the Sturmian basis of the group.
The Sturmian basis of the algebra  ${\rm su}(1,1)$ for the standard (non-relativistic) Kepler--Coulomb potential in 3D were reported in Ref. \cite{gerry } and generalized in Ref. \cite{gur} to D dimensions. The ${\rm su}(1,1)$ Lie algebra generators of Ref. \cite{gur} for non-relativistic Coulomb potential in D dimensions are given by
\begin{align}\label{g1}
\mathcal{K}_0&=\frac{1}{2}\left(-r\frac{d^2}{d r^2}-(D-1)\frac{d}{d r}-\frac{L(L+D-2)}{r}+r\right),\\\label{g2}
\mathcal{K}_1&=\frac{1}{2}\left(-r\frac{d^2}{d r^2}-(D-1)\frac{d}{d r}+\frac{L(L+D-2)}{r}-r\right),\\\label{g3}
\mathcal{K}_2&=-i\left(r\frac{d}{d r}+\frac{D-1}{2}\right),
\end{align}
and have the Sturmian basis
\begin{equation}
S_{N,L}(r)=2\left[\frac{\Gamma(N+1)}{\Gamma\left(N+2L+D-1\right)}\right]^{1/2}(2 r)^{ L} e^{-r}L_{N}^{2L+D-2}\left(2r\right). \label{sg}
\end{equation}
The formal comparison of our  ${\rm su}(1,1)$  operators, equations  (\ref{A1})-(\ref{A3}),  with those of equations (\ref{g1})-(\ref{g3}) leads us to find
\begin{align}
D&=2\mu_1+2\mu_2+2,\\
L&=-\mu_1-\mu_2+\sqrt{(2\ell+\mu_1+\mu_2 )^2-\gamma^2}.
\end{align}
In this way, we find that the Sturmian basis of the DKG equation for the Coulomb potential is given by
\begin{align}
&\widetilde{R}_{n,\ell}(\rho)=\widetilde{C}_{n,\ell}\,(2\rho)^{ -\mu_1-\mu_2+\sqrt{(2\ell+\mu_1+\mu_2 )^2-\gamma^2}} e^{-\rho}L_{n}^{2 \sqrt{(2\ell+\mu_1+\mu_2 )^2-\gamma^2}}\left(2\rho\right), \label{sturmian}\\
&\widetilde{C}_{n,\ell}=2\left[\frac{\Gamma(n+1)}{\Gamma\left(n+2\sqrt{(2\ell+\mu_1+\mu_2 )^2-\gamma^2}+1 \right)}\right]^{1/2}.
\end{align}
By using the scaling operator equation  \cite{KTH}
\begin{equation}\label{scall}
e^{i\theta{A_2}}f(\rho)=e^{\theta}f(e^{\theta}\rho),
\end{equation}
for $f(\rho)$ an arbitrary spherically symmetric function, and $\theta$ the scaling parameter, we obtain the physical radial states from the Sturmian basis states. Applying this property to the Sturmian basis (\ref{sturmian}), we  obtain the radial physical states $R_{n,\ell}(\rho)$ of equation (\ref{eigen}) for the Dunkl--Klein--Gordon equation with Coulomb potential
\begin{equation}
R_{n,\ell}(\rho)=C\,(2\sqrt{\alpha}\rho)^{-\mu_1-\mu_2+\sqrt{(2\ell+\mu_1+\mu_2 )^2-\gamma^2}} e^{-\sqrt{\alpha}\rho}L_{n}^{2 \sqrt{(2\ell+\mu_1+\mu_2 )^2-\gamma^2}}\left(2\sqrt{\alpha}\rho\right),\label{Rfin}
\end{equation}
where the scaling parameter  is obtained from equations (\ref{def}) and (\ref{eigen})  as  $e^{\theta}=\sqrt{\alpha}=\sqrt{\frac{m^2c^4-E^2}{\hbar^2c^2}}$, and $C$ is a constant, which is explicitly calculated in the next subsection.

\subsection{Analytical solution of the DKG--Coulomb problem}

In order to obtain the analytical solution of the DKG equation for the Coulomb potential, we study the equation
\begin{equation}
\left(\frac{d^2}{d \rho^2}+\frac{A}{\rho}\frac{d}{d \rho}-\frac{B}{\rho^2}+\frac{\lambda}{\rho}-\alpha\right)f(\rho)=0,
\end{equation}
which is more general than the radial equation (\ref{rad}) of our problem. Now, we apply the change of variable $\rho=\frac{r}{2\sqrt{\alpha}}$ to obtain the differential equation
\begin{equation}
\left(r\frac{d^2}{d r^2}+A\frac{d}{d r}-\frac{B}{r}+\frac{\lambda}{2\sqrt{\alpha}}-\frac{r}{4}\right)f(r)=0.\label{gen}
\end{equation}
On the other hand, it is known that the differential equation
\begin{equation}
xu''+(\beta+1-2\nu)u'+\left(n+\frac{\beta+1}{2}+\frac{\nu(\nu-\beta)}{x}-\frac{x}{4}\right)u=0,\label{generaleq}
\end{equation}
has as solution the function \cite{LEB}
\begin{equation}
u(x)=Ce^{-\frac{x}{2}}x^\nu L_n^\beta(x),\hspace{5ex} n=0,1,2,...,\label{colsol}
\end{equation}
where $C$ is an arbitrary constant and $L_n^\beta(x)$ are the generalized Laguerre polynomials. Comparing equations (\ref{gen}) and (\ref{generaleq}), we identify $x$ with $r$, and $u$ with $f(r)$, and obtain the following set of equations
\begin{equation}
\beta-2\nu+1=A,\hspace{7ex}
\nu(\nu-\beta)=-B,\hspace{7ex}
n+\frac{\beta+1}{2}=\frac{\lambda}{2\sqrt{\alpha}}.  \label{rest}
\end{equation}
From the first two equations we find
\begin{eqnarray}
&&\nu=\frac{1}{2}\left(1-A+\sqrt{A^2-2A+4B+1}\right),\\ \label{lagc1}
&&\beta=\sqrt{A^2-2A+4B+1}.\label{lagc2}
\end{eqnarray}
By considering that $\lambda=\frac{2E\gamma}{\hbar c}$ and $\alpha=\frac{m^2c^2-E^2}{\hbar^2c^2}$, from the last equation of
(\ref{rest}), we obtain that the energy spectrum of the equation (\ref{gen}) is
\begin{equation}
E=\pm mc^2\left(1+\frac{\gamma^2}{n+\frac{1}{2}\sqrt{A^2-2A+4B+1}+\frac{1}{2}}\right)^{-\frac{1}{2}}.\label{genesp}
\end{equation}
In particular, for the DKG equation for the Coulomb potential,  equation (\ref{rad}), the parameters $A$ and $B$ are
\begin{equation}
A=1+2\mu_1+2\mu_2,\hspace{10ex}B=s^2-\gamma^2=4\ell(\ell+\mu_1+\mu_2)-\gamma^2. \label{cpar}
\end{equation}
Substituting these values of $A$ and $B$ into equation (\ref{genesp}) we obtain that this energy spectrum is equal to the spectrum of equation (\ref{lieesp}) obtained by algebraic methods. Hence, the values of the parameter $A$ and $B$ of equation (\ref{cpar}) lead to the parameters of the solutions (\ref{colsol}) to take the values
\begin{eqnarray}
&&\beta=2\sqrt{(2\ell+\mu_1+\mu_2)^2-\gamma^2},\label{beta}\\
&&\nu=-\mu_1-\mu_2+\sqrt{(2\ell+\mu_1+\mu_2)^2-\gamma^2}.\label{nu}
\end{eqnarray}
The constant $C$ of equation (\ref{colsol}) for our problem can be computed from the expression
\begin{equation}
\int_0^{\infty}e^{-x}x^{\alpha+1}\left[L_{n}^{\alpha}(x)\right]^2dx=\frac{\Gamma(n+\alpha+1)}{n!}(2n+\alpha+1).\label{norm+1}
\end{equation}
Thus, the normalization constant $C$, which depends on $n$, $\ell$, $\mu_1$, and $\mu_2$, is given by
\begin{equation}
C=\sqrt{\frac{n!}{\Gamma(n+2\sqrt{(2\ell+\mu_1+\mu_2)^2-\gamma^2}+1)(2n+2\sqrt{(2\ell+\mu_1+\mu_2)^2-\gamma^2}+1)}},
\end{equation}
where we have used Eqns. (\ref{beta}) and (\ref{nu}). Therefore, the energy spectrum and radial eigenfunctions $(\rho=\frac{r}{2\sqrt{\alpha}})$ of the DKG equation for the 2D Coulomb potential are explicitly given by
\begin{equation}
E=\pm mc^2\left(1+\frac{\gamma^2}{(n+\sqrt{(2\ell+\mu_1+\mu_2 )^2-\gamma^2}+\frac{1}{2})^2} \right)^{-\frac{1}{2}},
\end{equation}
and
\begin{eqnarray}
&&R_{n\ell}(\rho)=\sqrt{\frac{n!}{\Gamma(n+2\sqrt{(2\ell+\mu_1+\mu_2)^2-\gamma^2}+1)(2n+2\sqrt{(2\ell+\mu_1+\mu_2)^2-\gamma^2}+1)}}\nonumber\\
&&\hspace{10ex}\times (2\sqrt{\alpha}{\rho})^{-\mu_1-\mu_2+\sqrt{(2\ell+\mu_1+\mu_2)^2-\gamma^2}}e^{-\sqrt{\alpha}\rho} L_n^{2\sqrt{(2\ell+\mu_1+\mu_2)^2-\gamma^2}}(2\sqrt{\alpha}\rho),\label{colsolfin}
\end{eqnarray}
where $n=0,1,2,...$.

It can be shown that these radial functions are normalized according to the following expression \cite{GEN4}
\begin{equation}
\int_0^{\infty}R_{n\ell}(\rho)R_{n'\ell}(\rho){\rho}^{1+2\mu_1+2\mu_2}d\rho=\delta_{nn'}.\label{normgen}
\end{equation}
Therefore, the wavefunctions $\Psi_{C}=R_{n\ell}(\rho)\Phi_\ell^{(e_1,e_2)}(\phi)$ for the DKG Coulomb potential are orthogonal against the scalar product
\begin{equation}
\langle f,g\rangle=\int_0^{\infty}\int_0^{2\pi}f^{*}(\rho,\phi)g(\rho,\phi)|\rho\cos{\phi}|^{2\mu_1}|\rho\sin{\phi}|^{2\mu_2}\rho d\rho d\phi.\label{prodgen}
\end{equation}
We emphasize that the energy spectrum and the physical states obtained analytically are in total agreement with those found in subsection 2.1 in an algebraic way.

\section{The DKG equation for the 2D Klein--Gordon oscillator}

The standard Klein--Gordon oscillator equation for stationary states in 2D  is \cite{bruce,mex,roa,boumali,bras}
\begin{equation}
(E^2-m^2c^4)\Psi_{O}=c^2\left({\bf P}+im\omega\rho \hat \rho\right)\cdot \left({\bf P}-im\omega\rho \hat \rho\right)\Psi_{O},
\end{equation}
where $\rho=\sqrt{x^2+y^2}$ and $\hat \rho$ is a unitary radial vector. By changing the standard partial derivative by the Dunkl derivative, and using the last commutation relations of equation (\ref{pro2}), we obtain
\begin{equation}
H_O\Psi_{O}\equiv\left(-\hbar^2(D_1^2+D_2^2)+m^2\omega^2\rho^2+2\hbar m\omega(1+\mu_1R_1+\mu_2R_2)+m^2c^2-\left(\frac{E}{c}\right)^2\right)\Psi_{O}=0,\hspace{2ex}\label{DKGO}
\end{equation}
where we have defined the pseudo-Hamiltonian $H_O$. Using the results (\ref{res1})-(\ref{res4}) it is immediate to show that the $z$-component of the angular momentum $L_z=-i\hbar(xD_2-yD_1)$ commutes with $H_O$:
\begin{equation}
[L_z,H_O]=0.
\end{equation}

If we define $r=\sqrt{\frac{m\omega}{\hbar}}\rho$, use the Dunkl Laplacian in polar coordinates (\ref{laplapol}), and that the eigenvalues of the angular operator $B_\phi$ are $\frac{s^2}{2}=2\ell(\ell+\mu_1+\mu_2)$, we obtain the DKG radial equation for the 2D Klein--Gordon oscillator
\begin{equation}
\left(-\frac{d^2}{d r^2}-\frac{1+2\mu_1+2\mu_2}{r}\frac{d}{d r}+\frac{s^2}{r^2}+r^2\right)R(r)=\epsilon R(r)\label{rado}
\end{equation}
where we have defined $\epsilon$ as
\begin{equation}
\epsilon\equiv \left(\frac{E^2-m^2c^4}{m\omega\hbar c^2}\right)-2(1+\mu_1R_1+\mu_2R_2).\label{espectrito}
\end{equation}
In what follows, we solve this equation using algebraic and analytical approaches similar to those introduced in section 2.

\subsection{Algebraic approach for the DKG-oscillator}

To solve equation (\ref{rado}) by algebraic methods, we use the experience gained in solving algebraically the Shr\"odinger equation for the 2D harmonic oscillator \cite{nos1}. Thus we find the set of operators
\begin{eqnarray}
&&O_0=\frac{1}{4}\left(-\frac{d^2}{dr^2}-\frac{1+2\mu_1+2\mu_2}{r}\frac{d}{dr}+\frac{s^2}{r^2}+r^2\right),\\\label{o1}
&&O_+=\frac{1}{2}\left(-r\frac{d}{dr}+r^2 -(1+\mu_1+\mu_2)-2O_0\right),\\
&&O_-=\frac{1}{2}\left(r\frac{d}{dr}+r^2 +(1+\mu_1+\mu_2)-2O_0\right),\\\label{o3}
\end{eqnarray}
which span the Lie algebra ${\rm su}(1,1)$ (equation (\ref{com})). A direct calculation leads us to show that the Casimir operator is given by
\begin{equation}
O^2=\frac{s^2+(\mu_1+\mu_2)^2-1}{4}.
\end{equation}
According to the ${\rm su}(1,1)$ representation theory, this value must be equal to $k(k-1)$. From this, we obtain that for the discrete series $k>0$,
\begin{equation}
k=\frac{1}{2}\left(1+\sqrt{s^2+(\mu_1+\mu_2)^2}\right)=\ell+\frac{1+\mu_1+\mu_2}{2},
\end{equation}
where we have used that $s^2=4\ell(\ell+\mu_1+\mu_2)$. By writing the left hand side of Eqn. (\ref{rado}) in terms of the $O_0$ operator, and using Eqn. (\ref{k0n}), we have
\begin{equation}
O_0R(r)=(n+k)R(r)=\frac{1}{4}\epsilon R(r),
 \end{equation}
From the second equality and the definition of $\epsilon$ (equation (\ref{espectrito})), we get that the energy spectrum is given by
\begin{equation}
E=\pm mc^2\left(1+\frac{4\hbar \omega}{mc^2}\left(n+\ell +1+\frac{\mu_1(1+R_1)+\mu_2(1+R_2)}{2}\right)\right)^\frac{1}{2}.\label{especosc}
\end{equation}
It is convenient to remember that the eigenvalues of the operators $(R_1,R_2)$ are given by $(1-2e_1,1-2e_2)$, from which we obtain Table \ref{l table}.
\begin{table}[ht]
\begin{center}
\begin{tabular}{| r | l | c |}\hline
$(e_1,e_2) $& $(R_1,R_2)$ & $s^2=4\ell(\ell+\mu_1+\mu_2)$ \\ \hline
$(0,0)$ & $(1,1)$ &$\ell$  non-negative integer\\
$(1,1)$ & $(-1,-1)$ & $\ell$ non-negative integer \\
$(0,1)$ & $(1,-1)$& $\ell$ positive half integer \\
$(1,0)$& $(-1,1)$&$\ell$ positive half integer\\ \hline
\end{tabular}
\caption{Possible values of $\ell$ in terms of the eigenvalues of $(R_1,R_2)$}
\label{l table}
\end{center}
\end{table}
According to the information presented in Table \ref{l table} we have the following particular cases for the spectrum (\ref{especosc}): \\
1) If $\ell$ is a non-negative integer, and the eigenvalues of $(R_1,R_2)$ are $(1,1)$, then
\begin{equation}
E=\pm mc^2\left(1+\frac{4\hbar \omega}{mc^2}\left(n+\ell+1 +\mu_1+\mu_2\right)\right)^\frac{1}{2},
\end{equation}
2) If $\ell$ is a non-negative integer, and the eigenvalues of $(R_1,R_2)$ are $(-1,-1)$,
\begin{equation}
E=\pm mc^2\left(1+\frac{4\hbar \omega}{mc^2}\left(n+\ell +1\right)\right)^\frac{1}{2},
\end{equation}
3) If $\ell$ is a positive half integer, and the eigenvalues of $(R_1,R_2)$ are $(1,-1)$,
\begin{equation}
E=\pm mc^2\left(1+\frac{4\hbar \omega}{mc^2}\left(n+\ell +\mu_1\right)\right)^\frac{1}{2},
\end{equation}
4) If $\ell$ is a positive half integer, and the eigenvalues of $(R_1,R_2)$ are $(-1,1)$,
\begin{equation}
E=\pm mc^2\left(1+\frac{4\hbar \omega}{mc^2}\left(n+\ell +\mu_2\right)\right)^\frac{1}{2}.
\end{equation}
We emphasize that in this case it was not necessary to use the tilting technique to find the energy spectrum, as in the case of the Coulomb potential. However, in this problem the energy spectrum explicitly depends on the eigenvalues of the reflection operators $(R_1,R_2)$.

The ${\rm su}(1,1)$ Lie algebra to study the  non-relativistic harmonic oscillator in D dimensions were reported in Ref. \cite{gur}, and are given by
\begin{eqnarray}
&&\mathcal{O}_0=\frac{1}{4}\left(-\frac{d^2}{dr^2}-\frac{D-1}{r}\frac{d}{dr}+\frac{L(L+D-2)}{r^2}+r^2\right),\label{oo1}\\
&&\mathcal{O}_+=\frac{1}{2}\left(-r\frac{d}{dr}+r^2 -\frac{D}{2}-2\mathcal{O}_0\right),\\
&&\mathcal{O}_-=\frac{1}{2}\left(r\frac{d}{dr}+r^2+\frac{D}{2} -2\mathcal{O}_0\right),\label{oo3}
\end{eqnarray}
with its corresponding Sturmian basis
\begin{equation}
R_{N,L}(r)=\left[\frac{2\Gamma(N+1)}{\Gamma\left(N+2L+\frac{D}{2}\right)}\right]^{1/2}r^{L} e^{-\frac{r^2}{2}}L_{N}^{L+\frac{D-2}{2}}\left(r^2\right). \label{sto}
\end{equation}
The formal comparison of our ${\rm su}(1,1)$ Lie algebra generators, equations (\ref{o1})-(\ref{o3}), with the operators (\ref{oo1})-(\ref{oo3}) leads to
\begin{equation}
L=2\ell, \hspace{10ex}D=2\mu_1+2\mu_2+2.
\end{equation}
Thus, for our case of the DKG oscillator, the Sturmian basis results to be
\begin{equation}
R_{n,\ell}(r)=C_0\,r^{ 2\ell} e^{-\frac{r^2}{2}}L_{n}^{2\ell+\mu_1+\mu_2 }\left(2r\right), \label{sturmian2}
\end{equation}
where $C_0$ is a constant that is found in the next subsection.

We note that the algebraic solution in this case was found in a more direct way than the algebraic solution of the DKG for the Coulomb potential. This is because, in the present case, the scaling factor is energy-independent, and therefore, the physical functions and the Sturmian basis result to be the same.

\subsection{Analytical solution of the DKG-oscillator}

Now, we proceed to solve the DKG equation for the KG oscillator, Eqn. (\ref{rado}) in an analytical form. To do this, we consider the more general differential equation
\begin{equation}
\left(-\frac{d^2}{d r^2}-\frac{\mathcal A}{r}\frac{d}{d r}+\frac{\mathcal B}{r^2}+r^2\right)R(r)={\mathcal E}R(r).\label{rado2}
\end{equation}
Setting $R(r)=r^{-\frac{1}{2}\mathcal A} G(r)$, and rearranging we obtain
\begin{equation}
\left(\frac{d^2}{d r^2}+{\mathcal E}- r^2+\frac{\frac{1}{2}{\mathcal A}-\frac{1}{4}{\mathcal A}^2-{\mathcal B}}{r^2}\right)G(r)=0.\label{nosotros}
\end{equation}
This equation has the same form of the differential equation
 \begin{equation}
u''+\left(4n+2\alpha+2-x^2+\frac{\frac{1}{4}-\alpha^2}{x^2}\right)u=0,\label{levedev}
\end{equation}
which has as solution the function \cite{LEB}
\begin{equation}
u(x)=C_0e^{-\frac{x^2}{2}}x^{\alpha+\frac{1}{2}}L_n^\alpha(x^2)\hspace{5ex} n=0,1,2,...
\end{equation}
where again, $C_0$ is an arbitrary constant and $L_n^\alpha(x^2)$ are the generalized Laguerre polynomials. Thus, the comparison between equations (\ref{nosotros}) and (\ref{levedev}) leads to
\begin{equation}
\alpha=\frac{1}{2}\sqrt{({\mathcal A}-1)^2+4{\mathcal B}}\hspace{7ex} {\mathcal{E}}=4n+2+\sqrt{({\mathcal A}-1)^2+4{\mathcal B}}\label{uno}
\end{equation}
Equations (\ref{rado}) and (\ref{rado2}) coincide with the identifications
\begin{equation}
 {\mathcal E}=\epsilon \hspace{10ex}{\mathcal B}=s^2=4\ell(\ell+\mu_1+\mu_2)\hspace{10ex} {\mathcal A}=1+2\mu_1+2\mu_2.\label{dos}
\end{equation}
With these particular values it results that $\alpha=2\ell+\mu_1+\mu_2$.
Using these expressions and Eqns. (\ref{espectrito}), (\ref{uno}) and (\ref{dos}), we immediately show that the energy spectrum is
\begin{equation}
E=\pm mc^2\left(1+\frac{4\hbar \omega}{mc^2}\left(n+\ell +1+\frac{\mu_1(1+R_1)+\mu_2(1+R_2)}{2}\right)\right)^\frac{1}{2},
\end{equation}
which is the same as that  given by the equation (\ref{especosc}). Hence, when $\ell$ is a non-negative integer, and the eigenvalues of $(R_1,R_2)$ are $(1,1)$ or $(-1,-1)$, we recover  the case 1) and 2) above, respectively. When $\ell$ is a positive half integer, and the eigenvalues of $(R_1,R_2)$ are $(1,-1)$ or $(-1,1)$ the cases 3) and 4) holds, respectively.

For this problem the normalization constant $C_0$ can be determined by using $\alpha=2\ell+\mu_1+\mu_2$ and the orthogonality of the Laguerre polynomials
\begin{equation}
\int_0^{\infty}e^{-x}x^{\alpha}\left[L_{n}^{\alpha}(x)\right]^2dx=\frac{\Gamma(n+\alpha+1)}{n!}.\label{norm}
\end{equation}
It results to be
\begin{equation}
C_0=\sqrt{\frac{2n!}{\Gamma(n+2\ell+\mu_1+\mu_2+1)}}.
\end{equation}
Consequently, the eigenfunctions of the DKG equation for the Klein--Gordon oscillator are explicitly given by
\begin{equation}
R_{n\ell}(r)_O=\sqrt{\frac{2n!}{\Gamma(n+2\ell+\mu_1+\mu_2+1)}}e^{-\frac{r^2}{2}}r^{2\ell}L_n^{2\ell+\mu_1+\mu_2}(r^2).
\end{equation}
These radial functions are also normalized according to the equation (\ref{normgen}) and therefore, the wavefunctions $\Psi_{O}=R_{n\ell}(r)_O\Phi_\ell^{(e_1,e_2)}(\phi)$ are orthogonal against the scalar product (\ref{prodgen}).
In this way we have computed the energy spectrum and eigenfunctions of the DKG oscillator and shown that the analytical and algebraic methods used are in complete agreement.

\section{Concluding remarks}

In the present paper, we constructed the DKG equation for a charged particle subject to a Coulomb potential. Then,
we have shown that the DKG equation for the Coulomb potential and the Klein--Gordon oscillator are exactly soluble in two different ways.
In the first way, we properly introduced for each problem a set of generators that span the Lie algebra ${\rm su}(1,1)$ and used their properties to obtain the energy spectrum and eigenfunctions.
In the second way, we solved the DKG equation for each problem analytically and show that the energy spectrum and eigenfunctions obtained are in full agreement with those obtained algebraically.

We notice that if $\mu_1=\mu_2=0$, the $s^2=4\ell^2$. In this case the angular part of the Laplacian is $B_\phi=-\frac{1}{2}\frac{\partial^2 }{\partial \phi^2}$, with eigenfunctions $e^{im\phi}$ and eigenvalues $\frac{m^2}{2}$. Since in general $B_\phi=\frac{s^2}{2}$, then when the Dunkl derivative parameters vanish, $s=m=2\ell$. With this identification, if we set the Dunkl parameters to vanish, $\mu_1=0$ and $\mu_2=0$, our results are in full agreement with those reported in Ref. \cite{dong} for the Klein--Gordon equation with a Coulomb potential in $D$ dimensions with $D=2$. Similarly, if we set $\mu_1=0$ and $\mu_2=0$ in our results obtained for the DKG oscillator, they are suitably reduced to those presented in reference \cite{Chargui} for the $D$-dimensional Klein--Gordon oscillator, with $D=2$ and $\beta=\beta'=0$, to obtain the standard Heisenberg algebra.

Thus, the DKG equation for the Coulomb potential and the Klein--Gordon oscillator are two other physical problems involving Dunkl operators which are exactly soluble.

\section*{Acknowledgments}
This work was partially supported by SNI-M\'exico, COFAA-IPN, EDI-IPN, EDD-IPN, and CGPI-IPN Project Number $20210734$.

\end{document}